\documentclass[a4paper, 10 pt, conference]{ieeeconf}
\IEEEoverridecommandlockouts
\usepackage{amsfonts,amsmath,amssymb} 

\usepackage{psfrag,color}
\usepackage{enumerate,cite,latexsym,graphicx}
\newtheorem{theorem}{Theorem}
\newtheorem{lemma}{Lemma}

\newtheorem{definition}{Definition}

\newtheorem{assumption}{Assumption}
\def\tr{\mathop{\rm Tr}\nolimits} 

\title{\LARGE \bf Robust Stability of Quantum Systems with a Nonlinear Coupling Operator
}

\author{Ian R.~Petersen, Valery Ugrinovskii and Matthew R~James  %
\thanks{This work was supported by the
Australian Research Council (ARC) and Air Force Office of Scientific
Research (AFOSR). This material is based on research sponsored by the
Air Force Research Laboratory, under agreement number
FA2386-09-1-4089.  The U.S. Government is authorized to reproduce and
distribute reprints for Governmental purposes notwithstanding any
copyright notation thereon.
The views and conclusions contained herein are those of the authors
and should not be interpreted as necessarily representing the official
policies or endorsements, either expressed or implied, of the Air
Force Research Laboratory or the U.S. Government. }%
\thanks{Ian R. Petersen and Valery Ugrinovskii are with the School of  Engineering and Information Technology, 
        University of New South Wales at the Australian Defence Force Academy, Canberra ACT 2600, Australia.
         {\tt\small \{i.r.petersen,v.ugrinovskii\}@gmail.com} } 
\thanks{Matthew R. James is with the Research School of  Engineering,
College of Engineering and Computer Science, The Australian National University, Canberra, ACT 0200,
Australia. Email: Matthew.James@anu.edu.au.}
}%

\begin{document}

\maketitle
\thispagestyle{empty}
\pagestyle{empty}

\begin{abstract}
This paper considers the problem of robust stability for a class of
uncertain quantum systems subject to unknown perturbations in the
system coupling operator. A general stability result is given for a
 class of perturbations to the system coupling operator. Then, the
special case of a nominal linear quantum system is considered with
 non-linear perturbations to the system
coupling operator. In this case, a robust stability condition is given in terms of a scaled strict bounded real condition. 
\end{abstract}

\section{Introduction} \label{sec:intro}
An important concept in modern control theory is the notion of
robust or absolute stability for uncertain nonlinear systems in the form of a
Lur'e system with an uncertain nonlinear block which satisfies a sector bound
condition; e.g., see \cite{KHA02}. This enables a frequency domain
condition for robust stability to be given. This
characterization of robust stability enables robust feedback
controller synthesis to be carried out using $H^\infty$ control
theory; e.g., see \cite{ZDG96}. In a recent paper \cite{PUJ1a}, classical results on  robust  stability were extended to the case of nonlinear quantum systems with non-quadratic perturbations to the system Hamiltonian. 
The aim of this paper is to  extend
classical results on  robust  stability to the case of
nonlinear quantum systems with nonlinear perturbations to the system coupling operator. 

In recent years, a number of papers have considered the feedback
control of systems whose dynamics are governed by the laws of quantum
mechanics rather than classical mechanics; e.g., see
\cite{YK03A,YK03B,YAM06,JNP1,NJP1,GGY08,MaP3,MaP4,YNJP1,GJ09,GJN10,WM10,PET10Ba}. In
particular, the papers \cite{GJ09,JG10} consider a framework of
quantum systems defined in terms of a triple $(S,L,H)$ where $S$ is a
scattering matrix, $L$ is a vector of coupling operators and $H$ is a
Hamiltonian operator. The paper \cite{JG10} then introduces notions of
dissipativity and stability for this class of quantum systems. In this
paper, we build on the results of \cite{JG10} to obtain robust 
stability results for uncertain quantum systems in which the quantum system coupling operator is
decomposed as $L =L_1+L_2$ where $L_1$ is a known nominal coupling operator
and $L_2$ is a perturbation coupling operator, which is contained in a
specified set of coupling operators $\mathcal{W}$. 

For this general class of uncertain quantum systems, a general stability
result is obtained. The paper then considers the case in which the
nominal quantum system $(S,L_1,H)$ is a linear quantum system in which the Hamiltonian $H$ is a quadratic function of annihilation and
creation operators and the coupling operator $L_1$ is a linear
function of annihilation and creation operators; e.g., see
\cite{JNP1,NJP1,MaP3,MaP4,PET10Ba}. In this special case, a robust stability
result is obtained in terms of a scaled frequency domain condition. 

The remainder of the paper proceeds as follows. In Section
\ref{sec:systems}, we define the general class of uncertain quantum
systems under consideration.  In Section
\ref{sec:nonquadratic}, we consider a  special class of  non-linear perturbation coupling operators. In Section
\ref{sec:linear}, we specialize to the case of a linear nominal
quantum systems and obtain a robust stability result for this case
in which the stability condition is given in terms of a strict bounded
real condition dependent on three scaling parameters. In Section  \ref{sec:conclusions},
we present some conclusions. 

\section{Quantum Systems} \label{sec:systems}
We consider  open quantum systems defined by  parameters $(S,L,H)$ where $L = L_1+L_2$; e.g., see \cite{GJ09,JG10}.  The corresponding generator for this quantum system is given by 
\begin{equation}
\label{generator}
\mathcal{G}(X) = -i[X,H] + \mathcal{L}_L(X)
\end{equation}
where $ \mathcal{L}_L(X) = \frac{1}{2}L^*[X,L]+\frac{1}{2}[L^*,X]L$. Here, $[X,H] = XH-HX$ denotes the commutator between two operators and the notation $^*$ denotes the adjoint  of an  operator. Also, $H$ is a self-adjoint operator on the underlying Hilbert space referred to as the system Hamiltonian.  $L_1$ is the nominal system coupling operator and  $L_2$ is referred to as the perturbation coupling operator.  Also, $S$ is a unitary matrix referred to as the scattering matrix. Throughout this paper, we will assume that $S=I$.  The triple $(S,L,H)$, along with the corresponding generator define the Heisenberg evolution $X(t)$ of an operator $X$ according to a quantum stochastic differential equation
\begin{eqnarray*}
dX &=&\left(\mathcal{L}_L(X)-i[X,H]\right)dt\\
&&+dA^*S^\dagger[X,L]+[L^*,X]SdA\\
&&+\mbox{tr}\left[\left(S^\dagger XS-X\right)d\Lambda^T\right]; 
\end{eqnarray*}
e.g., see \cite{JG10}. Here, in the case of operators, the notation $^\dagger$ denotes the adjoint transpose of a vector or matrix of operators; see also \cite{JG10} for a definition of the quantities $dA$, $dA^*$, and $d\Lambda$ which will not be further considered in this paper. Also, in the case of standard matrices, the notation $^\dagger$ refers to the complex conjugate transpose of a matrix. 

The problem under consideration involves establishing robust stability
properties for an uncertain open quantum system for the case in which the perturbation coupling operator is contained in a given set $\mathcal{W}$. 
The main robust stability results presented in this paper will build on the following result from \cite{JG10}. 
\begin{lemma}[See Lemma 3.4 of \cite{JG10}.]
\label{L0}
Consider an open quantum system defined by $(S,L,H)$ and suppose there exists a non-negative self-adjoint operator $V$ on the underlying Hilbert space such that
\begin{equation}
\label{lyap_ineq}
\mathcal{G}(V) + cV \leq \lambda
\end{equation}
where $c > 0$ and $\lambda$ are real numbers. Then for any plant state, we have
\[
\left<V(t)\right> \leq e^{-ct}\left<V\right> + \frac{\lambda}{c},~~\forall t \geq 0.
\]
Here $V(t)$ denotes the Heisenberg evolution of the operator $V$ and $\left<\cdot\right>$ denotes quantum expectation; e.g., see \cite{JG10}.
\end{lemma}

\subsection{Commutator Decomposition}
Given a set of non-negative self-adjoint operators $\mathcal{P}$ and  real parameters $\gamma > 0$, $\delta_1 \geq 0$, $\delta_2 \geq 0$, $\delta_3 \geq 0$, we now define a particular set of perturbation coupling operators $\mathcal{W}_1$. This set $\mathcal{W}_1$ is defined in terms of the commutator decomposition
\begin{eqnarray}
\label{alt_comm_condition}
[V,L_2] &=& w_1[V,\zeta]-\frac{1}{2}w_2\left[\zeta,[V,\zeta]\right]
\end{eqnarray}
for  $V \in \mathcal{P}$ where $w_1$, $w_2$ and $\zeta$ are given scalar
operators.  We say $L_2 \in \mathcal{W}_1$ if the following
sector bound condition holds:
\begin{equation}
\label{sector2a}
L_2^* L_2 \leq \frac{1}{\gamma^2}\zeta^* \zeta + \delta_1
\end{equation}
and
\begin{equation}
\label{sector2b}
w_1^* w_1 \leq  \delta_2,
\end{equation}
\begin{equation}
\label{sector2c}
w_2^* w_2 \leq \delta_3.
\end{equation}
Here, we use the  convention  that for operator inequalities, terms consisting of real constants are interpreted as that constant multiplying the identity operator. 

Then, we define
\begin{equation}
\label{W1}
\mathcal{W}_1 = \left\{\begin{array}{l}L_2: \exists w_1,~w_2,~\zeta \mbox{
      such that 
(\ref{sector2a}), (\ref{sector2b}) and (\ref{sector2c}) } \\
\mbox{ are satisfied and (\ref{alt_comm_condition}) is satisfied } \forall V \in \mathcal{P}\end{array}\right\}.
\end{equation}
Using this definition, we obtain the following theorem. 
\begin{theorem}
\label{T2}
Consider a set of non-negative self-adjoint operators $\mathcal{P}$
and an open quantum system $(S,L,H)$ where $L=L_1+L_2$ and $L_2 \in
\mathcal{W}_1$ defined in (\ref{W1}). If there exists a $V \in
\mathcal{P}$ and real constants  $c > 0$,
$\tilde \lambda \geq 0$, $\tau_1 > 0,~\tau_2 > 0,\ldots,\tau_5>0$  such that 
$\mu = -\frac{1}{2}\left[\zeta,[V,\zeta]\right]$ is a constant and
\begin{eqnarray}
\label{dissip1a}
&&-i[V,H]+ \mathcal{L}_{L_1}(V)
+\left(\frac{\tau_1^2}{2}+\frac{\tau_2^2}{2}\right)L_1^*L_1\nonumber \\
&&+\left(\frac{\delta_2}{2\tau_1^2}+\frac{\delta_2}{2\tau_4^2}\right)[V,\zeta]^*[V,\zeta]\nonumber \\
&&+\left(\frac{\tau_3^2}{2\gamma^2}+\frac{\tau_4^2}{2\gamma^2}+\frac{\tau_5^2}{2\gamma^2}\right)\zeta^* \zeta 
+\frac{[V,L_1]^*[V,L_1]}{2\tau_3^2}\nonumber \\
&&  
\nonumber \\
&&
+ cV \leq \tilde \lambda,
\nonumber \\
\end{eqnarray}
then 
\begin{eqnarray*}
\left<V(t)\right> &\leq& e^{-ct}\left<V\right>+ \nonumber \\
&&\frac{\tilde \lambda+\left(\frac{\delta_3}{2\tau_2^2}+\frac{\delta_3}{2\tau_5^2}\right)\mu^*\mu
+\left( \frac{\tau_3^2}{2} + \frac{\tau_4^2}{2}+ \frac{\tau_5^2}{2}\right)\delta_1}{c}
\end{eqnarray*}
for all $t \geq 0$.
\end{theorem}

\noindent
{\em Proof:}
Let $V \in \mathcal{P}$ be given and consider $\mathcal{G}(V)$ defined in (\ref{generator}). Then 
using (\ref{alt_comm_condition}) and the fact that $V$ is self-adjoint, 
\begin{eqnarray}
\label{ineq1a}
\lefteqn{\mathcal{G}(V)}\nonumber \\
 &=& -i[V,H]+\frac{1}{2}\left(L_1+L_2\right)^*[V,L_1+L_2] \nonumber \\
&&+\frac{1}{2}[\left(L_1+L_2\right)^*,V]\left(L_1+L_2\right) \nonumber \\
&=& -i[V,H]\nonumber \\
&&+\frac{1}{2}\left(L_1^*+L_2^*\right)
\left([V,L_1]+w_1[V,\zeta]+\mu w_2\right)\nonumber \\
&&+\frac{1}{2}\left([V,L_1]^*+[V,\zeta]^*w_1^*+\mu^*w_2^*\right)
\left(L_1+L_2\right) \nonumber \\
&=&  -i[V,H]+\frac{1}{2}L_1^*[V,L_1]+\frac{1}{2}[V,L_1]^*L_1 \nonumber \\
&&+ \frac{1}{2}L_1^*w_1[V,\zeta]+ \frac{1}{2}[V,\zeta]^*w_1^*L_1\nonumber \\
&&+ \frac{1}{2}\mu L_1^*w_2+ \frac{1}{2}\mu^*w_2^*L_1\nonumber \\
&&+\frac{1}{2}L_2^*[V,L_1]+\frac{1}{2}[V,L_1]^*L_2\nonumber \\
&&+\frac{1}{2}L_2^*w_1[V,\zeta]+\frac{1}{2}[V,\zeta]^*w_1^*L_2\nonumber \\
&&+\frac{1}{2}\mu L_2^*w_2+\frac{1}{2}\mu^*w_2^*L_2.
\end{eqnarray}
Now 
\begin{eqnarray*}
0 &\leq& \left(\tau_1L_1^*-\frac{[V,\zeta]^*w_1^*}{\tau_1}\right)
\left(\tau_1L_1-w_1\frac{[V,\zeta]}{\tau_1}\right)\nonumber \\
&=& \tau_1^2L_1^*L_1 -L_1^*w_1[V,\zeta]-[V,\zeta]^*w_1^*L_1\nonumber \\
&&+\frac{[V,\zeta]^*w_1^*w_1[V,\zeta]}{\tau_1^2}
\end{eqnarray*}
and hence
\begin{eqnarray}
\label{ineq3a}
\lefteqn{L_1^*w_1[V,\zeta]+ [V,\zeta]^*w_1^*L_1}\nonumber \\
&\leq& \tau_1^2L_1^*L_1+\frac{[V,\zeta]^*w_1^*w_1[V,\zeta]}{\tau_1^2}\nonumber \\
&\leq& \tau_1^2L_1^*L_1+\frac{\delta_2[V,\zeta]^*[V,\zeta]}{\tau_1^2}
\end{eqnarray}
using (\ref{sector2b}).
Also, 
\begin{eqnarray*}
0 &\leq& \left(\tau_2L_1^*-\frac{\mu^*w_2^*}{\tau_2}\right)
\left(\tau_2L_1-\frac{\mu w_2}{\tau_2}\right)\nonumber \\
&=&  \tau_2^2L_1^*L_1 -\mu L_1^*w_2-\mu^*w_2^*L_1\nonumber \\
&&+\frac{\mu^*\mu w_2^*w_2}{\tau_2^2}
\end{eqnarray*}
and hence
\begin{eqnarray}
\label{ineq3b}
\mu L_1^*w_2+ \mu^*w_2^*L_1&\leq& \tau_2^2L_1^*L_1 +\frac{\mu^*\mu w_2^*w_2}{\tau_2^2}\nonumber \\
&\leq& \tau_2^2L_1^*L_1 +\frac{\delta_3\mu^*\mu}{\tau_2^2}
\end{eqnarray}
using (\ref{sector2c}). Also, 
\begin{eqnarray*}
0 &\leq& \left(\tau_3L_2^*-\frac{[V,L_1]^*}{\tau_3}\right)
\left(\tau_3L_2-\frac{[V,L_1]}{\tau_3}\right)\nonumber \\
&=&  \tau_3^2L_2^*L_2 -L_2^*[V,L_1]-[V,L_1]^*L_2\nonumber \\
&&+\frac{[V,L_1]^*[V,L_1]}{\tau_3^2}
\end{eqnarray*}
and hence
\begin{eqnarray}
\label{ineq3c}
\lefteqn{L_2^*[V,L_1]+[V,L_1]^*L_2}\nonumber \\
&\leq& \tau_3^2L_2^*L_2 +\frac{[V,L_1]^*[V,L_1]}{\tau_3^2}\nonumber \\
&\leq& \frac{\tau_3^2}{\gamma^2}\zeta^* \zeta + \tau_3^2\delta_1 +\frac{[V,L_1]^*[V,L_1]}{\tau_3^2}
\end{eqnarray}
using (\ref{sector2a}). Also, 
\begin{eqnarray*}
0 &\leq& \left(\tau_4L_2^*-\frac{[V,\zeta]^*w_1^*}{\tau_4}\right)
\left(\tau_4L_2-\frac{w_1[V,\zeta]}{\tau_4}\right)\nonumber \\
&=&  \tau_4^2L_2^*L_2 -L_2^*w_1[V,\zeta]-[V,\zeta]^*w_1^*L_2\nonumber \\
&&+\frac{[V,\zeta]^*w_1^*w_1[V,\zeta]}{\tau_4^2}
\end{eqnarray*}
and hence
\begin{eqnarray}
\label{ineq3d}
\lefteqn{L_2^*w_1[V,\zeta]+[V,\zeta]^*w_1^*L_2}\nonumber \\
&\leq& \tau_4^2L_2^*L_2 +\frac{[V,\zeta]^*w_1^*w_1[V,\zeta]}{\tau_4^2}\nonumber \\
&\leq& \frac{\tau_4^2}{\gamma^2}\zeta^* \zeta + \tau_4^2\delta_1 +\frac{\delta_2[V,\zeta]^*[V,\zeta]}{\tau_4^2}
\end{eqnarray}
using (\ref{sector2a}) and (\ref{sector2b}). Also, 
\begin{eqnarray*}
0 &\leq& \left(\tau_5L_2^*-\frac{\mu^*w_2^*}{\tau_5}\right)
\left(\tau_5L_2-\frac{\mu w_2}{\tau_5}\right)\nonumber \\
&=&  \tau_5^2L_2^*L_2 -\mu L_2^*w_2-\mu^*w_2^*L_2\nonumber \\
&&+\frac{\mu^*\mu w_2^*w_2}{\tau_5^2}
\end{eqnarray*}
and hence
\begin{eqnarray}
\label{ineq3e}
\lefteqn{\mu L_2^*w_2+\mu^*w_2^*L_2}\nonumber \\
&\leq& \tau_5^2L_2^*L_2 +\frac{\mu^*\mu w_2^*w_2}{\tau_5^2}\nonumber \\
&\leq& \frac{\tau_5^2}{\gamma^2}\zeta^* \zeta + \tau_5^2\delta_1 +\frac{\delta_3\mu^*\mu}{\tau_5^2}
\end{eqnarray}
using (\ref{sector2a}) and (\ref{sector2c}).

Substituting (\ref{ineq3a}), (\ref{ineq3b}), (\ref{ineq3c}), (\ref{ineq3d}) and  (\ref{ineq3e}) into (\ref{ineq1a}), it follows that
\begin{eqnarray}
\label{ineq2a}
\mathcal{G}(V) &\leq &  -i[V,H]+ \mathcal{L}_{L_1}(V)
+\frac{\tau_1^2}{2}L_1^*L_1+\frac{\delta_2[V,\zeta]^*[V,\zeta]}{2\tau_1^2}\nonumber \\
&&+\frac{\tau_2^2}{2}L_1^*L_1 +\frac{\delta_3\mu^*\mu}{2\tau_2^2}\nonumber \\
&&+\frac{\tau_3^2}{2\gamma^2}\zeta^* \zeta + \frac{\tau_3^2}{2}\delta_1 +\frac{[V,L_1]^*[V,L_1]}{2\tau_3^2}\nonumber \\
&&+\frac{\tau_4^2}{2\gamma^2}\zeta^* \zeta + \frac{\tau_4^2}{2}\delta_1 +\frac{\delta_2[V,\zeta]^*[V,\zeta]}{2\tau_4^2}
\nonumber \\
&&+\frac{\tau_5^2}{2\gamma^2}\zeta^* \zeta + \frac{\tau_5^2}{2}\delta_1 +\frac{\delta_3\mu^*\mu}{2\tau_5^2}.
\end{eqnarray}
 Then it follows from (\ref{dissip1a}) that 
\[
\mathcal{G}(V) + cV \leq \tilde \lambda+\left(\frac{\delta_3}{2\tau_2^2}+\frac{\delta_3}{2\tau_5^2}\right)\mu^*\mu
+\left( \frac{\tau_3^2}{2} + \frac{\tau_4^2}{2}+ \frac{\tau_5^2}{2}\right)\delta_1. 
\]
Then the result of the theorem follows from Lemma \ref{L0}.
\hfill $\Box$

\section{Non-linear  Perturbation Coupling Operators}
\label{sec:nonquadratic}
In this section, we define a  set of non-linear perturbation
coupling operators denoted $\mathcal{W}_2$.   For a given set of non-negative self-adjoint operators
$\mathcal{P}$ and real parameters $\gamma > 0$,  $\delta_1\geq 0$, $\delta_2\geq 0$, $\delta_3\geq 0$, 
consider perturbation coupling operators defined in terms of the following  power series (which is assumed to converge in some suitable sense)
\begin{equation}
\label{L2nonlin}
L_2 = f(\zeta) = \sum_{k=0}^\infty S_{k}\zeta^k = \sum_{k=0}^\infty  S_{k} L_{k}.
\end{equation}
Here  $\zeta$ is a scalar operator on the underlying Hilbert space and $L_{k} = \zeta^k$.

Also, we let 
\begin{equation}
\label{fdash}
f'(\zeta) = \sum_{k=1}^\infty k S_{k} \zeta^{k-1},
\end{equation}
\begin{equation}
\label{fddash}
f''(\zeta) = \sum_{k=1}^\infty k(k-1)S_{k} \zeta^{k-2}
\end{equation}
and consider the 
sector bound condition
\begin{equation}
\label{sector4a}
f(\zeta)^*f(\zeta)  \leq \frac{1}{\gamma^2}\zeta^* \zeta + \delta_1
\end{equation}
and the conditions
\begin{equation}
\label{sector4b}
f'(\zeta)^*f'(\zeta)  \leq \delta_2,
\end{equation}
\begin{equation}
\label{sector4c}
f''(\zeta)^*f''(\zeta) \leq  \delta_3.
\end{equation}
Then we define the set $\mathcal{W}_2$  as follows:
\begin{equation}
\label{W5}
\mathcal{W}_2 = \left\{\begin{array}{l}L_2 \mbox{ of the form
      (\ref{L2nonlin}) such that 
} \\
\mbox{ conditions (\ref{sector4a}), (\ref{sector4b}) and (\ref{sector4c}) are satisfied}\end{array}\right\}.
\end{equation}

In this section, the set of non-negative self-adjoint operators
$\mathcal{P}$ will be assumed to satisfy  the following  assumption:
\begin{assumption}
\label{A1}
Given any $V \in \mathcal{P}$, the quantity
\[
\mu = -\frac{1}{2}\left[\zeta,[V,\zeta]\right] = -\frac{1}{2}\zeta [V,\zeta]+\frac{1}{2}[V,\zeta]\zeta
\]
is a constant.
\end{assumption}

\begin{lemma}
\label{LB}
Suppose the set of self-adjoint operators $\mathcal{P}$ satisfies
Assumption \ref{A1}. Then
\[
\mathcal{W}_2 \subset \mathcal{W}_1.
\]
\end{lemma}

\noindent
{\em Proof:}
First, we note that given any $V \in \mathcal{P}$ and $k \geq 1$,
\begin{eqnarray}
\label{Vzetak}
 V\zeta  &=& [V,\zeta]+ \zeta V;\nonumber \\
\vdots && \nonumber \\
V\zeta^k &=& \sum_{n=1}^k \zeta^{n-1}[V,\zeta] \zeta^{k-n}+\zeta^k V.
\end{eqnarray}
Also for any $n \geq 1$ such that $n\leq k$,
\begin{eqnarray}
\label{Vzetak1}
[V,\zeta]\zeta  &=& \zeta[V,\zeta] + 2\mu; \nonumber \\
&\vdots & \nonumber \\
~[V,\zeta] \zeta^{k-n}
&=&\zeta^{k-n}[V,\zeta]
+ 2(k-n)\zeta^{k-n-1}\mu.
\end{eqnarray}
Therefore using (\ref{Vzetak}) and (\ref{Vzetak1}), it follows that
\begin{eqnarray*}
V\zeta^k &=&\sum_{n=1}^k  \zeta^{n-1}\zeta^{k-n}[V,\zeta]+ 2(k-n)\zeta^{n-1}\zeta^{k-n-1}\mu  \nonumber \\
&&+\zeta^k V\nonumber \\
&=& \sum_{n=1}^k  \zeta^{k-1}[V,\zeta]+ 2(k-n)\zeta^{k-2}\mu  +\zeta^k V\nonumber \\
&=&k \zeta^{k-1}[V,\zeta]+k(k-1)\zeta^{k-2}\mu+\zeta^k V,
\end{eqnarray*}
which holds for any $k \geq 0$. 

Now given any $L_2 \in \mathcal{W}_2$, $k \geq 0$ we have
\begin{eqnarray}
\label{VHkl}
[V,L_{k}] &=& k \zeta^{k-1}[V,\zeta]+k(k-1)\zeta^{k-2}\mu.
\end{eqnarray}
Therefore,
\begin{eqnarray}
\label{VH2}
[V,L_2] &=& \sum_{k=0}^\infty  S_{k} [V,L_{k}] \nonumber \\
&=& f'(\zeta)[V,\zeta]+  f''(\zeta)\mu.
\end{eqnarray}
Now letting 
\begin{equation}
\label{zw1w2}
w_1 = f'(\zeta),\mbox{ and }w_2=  f''(\zeta),
\end{equation}
 it follows that condition (\ref{alt_comm_condition}) is
 satisfied. Furthermore, conditions (\ref{sector2a}), (\ref{sector2b}), (\ref{sector2c})
 follow from conditions (\ref{sector4a}), (\ref{sector4b}), (\ref{sector4c})
 respectively. Hence, $L_2 \in
\mathcal{W}_1$. Since, $L_2 \in
\mathcal{W}_2$ was arbitrary, we must have $\mathcal{W}_2 \subset \mathcal{W}_1$.
\hfill $\Box$

\section{The  Case of a Linear Nominal System}
\label{sec:linear}
We now consider the  case in which the nominal quantum system corresponds to a linear quantum system; e.g., see \cite{JNP1,NJP1,MaP3,MaP4,PET10Ba}. In this case, we assume that $H$ is of the form 
\begin{equation}
\label{H1}
H = \frac{1}{2}\left[\begin{array}{cc}a^\dagger &
      a^T\end{array}\right]M
\left[\begin{array}{c}a \\ a^\#\end{array}\right]
\end{equation}
where $M \in \mathbb{C}^{2n\times 2n}$ is a Hermitian matrix of the
form
\[
M= \left[\begin{array}{cc}M_1 & M_2\\
M_2^\# &     M_1^\#\end{array}\right]
\]
and $M_1 = M_1^\dagger$, $M_2 = M_2^T$.  Here $a$ is a vector of annihilation
operators on the underlying Hilbert space and $a^\#$ is the
corresponding vector of creation operators.  In the case vectors of
operators, the notation $^\#$ refers to the vector of adjoint
operators and in the case of complex matrices, this notation refers to
the complex conjugate matrix. 

The annihilation and creation operators are assumed to satisfy the
canonical commutation relations:
\begin{eqnarray}
\label{CCR2}
\left[\left[\begin{array}{l}
      a\\a^\#\end{array}\right],\left[\begin{array}{l}
      a\\a^\#\end{array}\right]^\dagger\right]
&=&\left[\begin{array}{l} a\\a^\#\end{array}\right]
\left[\begin{array}{l} a\\a^\#\end{array}\right]^\dagger
\nonumber \\
&&- \left(\left[\begin{array}{l} a\\a^\#\end{array}\right]^\#
\left[\begin{array}{l} a\\a^\#\end{array}\right]^T\right)^T\nonumber \\
&=& J
\end{eqnarray}
where $J = \left[\begin{array}{cc}I & 0\\
0 & -I\end{array}\right]$; e.g., see \cite{GGY08,GJN10,PET10Ba}.

We also assume $L_1$ is of the form 
\begin{equation}
\label{L}
L_1 = \left[\begin{array}{cc}N_1 & N_2 \end{array}\right]
\left[\begin{array}{c}a \\ a^\#\end{array}\right] = \tilde N\left[\begin{array}{c}a \\ a^\#\end{array}\right]
\end{equation}
where $N_1 \in \mathbb{C}^{1\times n}$ and $N_2 \in
\mathbb{C}^{1\times n}$. Also, we write
\[
\left[\begin{array}{c}L_1 \\ L_1^*\end{array}\right] = N
\left[\begin{array}{c}a \\ a^\#\end{array}\right] =
\left[\begin{array}{cc}N_1 & N_2\\
N_2^\# &     N_1^\#\end{array}\right]
\left[\begin{array}{c}a \\ a^\#\end{array}\right].
\]

In addition we assume that $V$ is of the form 
\begin{equation}
\label{quadV}
V = \left[\begin{array}{cc}a^\dagger &
      a^T\end{array}\right]P
\left[\begin{array}{c}a \\ a^\#\end{array}\right]
\end{equation}
where $P \in \mathbb{C}^{2n\times 2n}$ is a positive-definite Hermitian matrix of the
form
\begin{equation}
\label{Pform}
P= \left[\begin{array}{cc}P_1 & P_2\\
P_2^\# &     P_1^\#\end{array}\right].
\end{equation}
 Hence, we consider the set of  non-negative self-adjoint operators
$\mathcal{P}_1$ defined as
\begin{equation}
\label{P1}
\mathcal{P}_1 = \left\{\begin{array}{l}V \mbox{ of the form
      (\ref{quadV}) such that $P > 0$ is a 
} \\
\mbox{  Hermitian matrix of the form (\ref{Pform})}\end{array}\right\}.
\end{equation}

In the linear case, we will consider a specific notion of robust mean square stability. 
\begin{definition}
\label{D1}
An uncertain open quantum system defined by  $(S,L,H)$ where $L=L_1+L_2$ with $L_1$ of the form (\ref{L}), $L_2 \in \mathcal{W}$, $\mathcal{W}$ is any given set, and $H$  of the form (\ref{H1}) is said to be {\em robustly mean square stable} if there exist constants $c_1 > 0$, $c_2 > 0$ and $c_3 \geq 0$ such that for any $L_2 \in \mathcal{W}$
\begin{eqnarray}
\label{ms_stable0}
\lefteqn{\left< \left[\begin{array}{c}a(t) \\ a^\#(t)\end{array}\right]^\dagger \left[\begin{array}{c}a(t) \\ a^\#(t)\end{array}\right] \right>}\nonumber \\
&\leq& c_1e^{-c_2t}\left< \left[\begin{array}{c}a \\ a^\#\end{array}\right]^\dagger \left[\begin{array}{c}a \\ a^\#\end{array}\right] \right>
+ c_3~~\forall t \geq 0.
\end{eqnarray}
Here $\left[\begin{array}{c}a(t) \\ a^\#(t)\end{array}\right]$ denotes the Heisenberg evolution of the vector of operators $\left[\begin{array}{c}a \\ a^\#\end{array}\right]$; e.g., see \cite{JG10}.
\end{definition}

In order to address the issue of robust mean square stability for the
uncertain linear quantum systems under consideration, we first require some algebraic identities.
\begin{lemma}
\label{L2}
Given $V \in \mathcal{P}_1$, $H$ defined as in (\ref{H1}) and $L_1$ defined as in (\ref{L}), then
\begin{eqnarray*}
\lefteqn{[V,H] =}\nonumber \\
&& \left[\left[\begin{array}{cc}a^\dagger &
      a^T\end{array}\right]P
\left[\begin{array}{c}a \\ a^\#\end{array}\right],\frac{1}{2}\left[\begin{array}{cc}a^\dagger &
      a^T\end{array}\right]M
\left[\begin{array}{c}a \\ a^\#\end{array}\right]\right] \nonumber \\
&=& \left[\begin{array}{c}a \\ a^\#\end{array}\right]^\dagger 
\left[
PJM - MJP 
\right] \left[\begin{array}{c}a \\ a^\#\end{array}\right].
\end{eqnarray*}

Also,
\begin{eqnarray*}
\lefteqn{\mathcal{L}_{L_1}(V) =} \nonumber \\
&& \frac{1}{2}L_1^\dagger[V,L_1]+\frac{1}{2}[L_1^\dagger,V]L_1 \nonumber \\
&=& \tr\left(PJN^\dagger\left[\begin{array}{cc}I & 0 \\ 0 & 0 \end{array}\right]NJ\right)
\nonumber \\&&
-\frac{1}{2}\left[\begin{array}{c}a \\ a^\#\end{array}\right]^\dagger
\left(N^\dagger J N JP+PJN^\dagger J N\right)
\left[\begin{array}{c}a \\ a^\#\end{array}\right].
\end{eqnarray*}

Furthermore, 
\[
\left[\left[\begin{array}{c}a \\ a^\#\end{array}\right],\left[\begin{array}{cc}a^\dagger &
      a^T\end{array}\right]P
\left[\begin{array}{c}a \\ a^\#\end{array}\right]\right] = 2JP\left[\begin{array}{c}a \\ a^\#\end{array}\right].
\]
\end{lemma}
{\em Proof:}
The proof of these identities follows via  straightforward but tedious
calculations using (\ref{CCR2}). \hfill $\Box$

We now specialize the results of Section \ref{sec:systems} to the case of a linear  nominal system with  $L_2 \in \mathcal{W}_2$ where $\mathcal{W}_2$ is defined as in Section \ref{sec:nonquadratic}. In this case, we define
\begin{eqnarray}
\label{z}
\zeta &=&   E_1a+E_2 a^\# \nonumber \\
&=& \left[\begin{array}{cc} E_1 & E_2 \end{array}\right]
\left[\begin{array}{c}a \\ a^\#\end{array}\right] =  \tilde E 
\left[\begin{array}{c}a \\ a^\#\end{array}\right]
\end{eqnarray}
where $\zeta$ is assumed to be a scalar operator. Then, we  show that a sufficient condition for robust mean square stability  is  the existence of constants $\tau_1 > 0$, $\tau_3 >  0$, and $\tau_4 > 0$ such that the following scaled 
  strict bounded real condition is satisfied:
\begin{enumerate}
\item
 The matrix 
\begin{equation}
\label{Hurwitz1}
F = -iJM-\frac{1}{2}JN^\dagger J N\mbox{ is Hurwitz;}
\end{equation}
\item
\begin{equation}
\label{Hinfbound1}
\left\|\bar C
\left(sI -F\right)^{-1}
\bar B \right\|_\infty < 1
\end{equation}
where
\begin{equation}
\label{barC}
\bar C = \left[\begin{array}{c}\frac{\sqrt{\tau_3^2+\tau_4^2}}{\gamma}\tilde E \\
\tau_1 \tilde N \end{array}\right]
\end{equation}
and
\begin{equation}
\label{barB}
\bar B = \left[\begin{array}{cc}\sqrt{\delta_2\left(\frac{1}{\tau_1^2}+\frac{1}{\tau_4^2}\right)}J\tilde E^\dagger & 
\frac{1}{\tau_3}J\tilde N^\dagger \end{array}\right].
\end{equation}
\end{enumerate}

This leads to the following theorem.

\begin{theorem}
\label{T4}
Consider an uncertain open quantum system defined by $(S,L,H)$  such that
$L=L_1+L_2$ where $L_1$ is of the form (\ref{L}), $H$ is of the
form (\ref{H1}) and $L_2 \in \mathcal{W}_2$. Furthermore, assume that there exists constants $\tau_1 > 0$, $\tau_3 >  0$, and $\tau_4 > 0$ such that 
the strict bounded real condition  (\ref{Hurwitz1}), (\ref{Hinfbound1})
is satisfied. Then the
uncertain quantum system is robustly mean square stable. 
\end{theorem}

In order to prove this theorem, we require the following lemma.
\begin{lemma}
\label{L4}
Given any $V \in \mathcal{P}_1$, then
\[
\mu = -\frac{1}{2}\left[\zeta,[V,\zeta]\right] = 
-\frac{1}{2} \tilde E \Sigma JPJ \tilde E^T.
\]
which is a constant. Here, $\Sigma = \left[\begin{array}{cc} 0 & I\\
I &0 \end{array}\right].
$
Hence, the set of operators $\mathcal{P}_1$ satisfies Assumption \ref{A1}. 
\end{lemma}
{\em Proof:}
The proof of this result follows via a straightforward but tedious
calculation using (\ref{CCR2}). \hfill $\Box$

\noindent
{\em Proof of Theorem \ref{T4}.}
If the conditions of the theorem are satisfied, then (\ref{Hinfbound1}) implies
\[
\left\|\frac{\bar C}{\sqrt 2}
\left(sI -F\right)^{-1}
\sqrt 2 \bar B \right\|_\infty < 1.
\]
Hence, 
it follows from the strict bounded real lemma that the matrix inequality 
\begin{equation*}
F^\dagger P + P F 
+2 P\bar B \bar B^\dagger P 
+ \frac{1}{2}\bar C^\dagger \bar C
 < 0.
\end{equation*}
will have a solution $P > 0$ of the form (\ref{Pform}); e.g., see \cite{ZDG96,MaP4}.  This matrix $P$ defines a corresponding operator $V \in \mathcal{P}_1$ as in (\ref{quadV}). Then using (\ref{barC}) and (\ref{barB}), it follows that we can write. 
\begin{eqnarray*}
\lefteqn{F^\dagger P + P F}\nonumber \\
&& +P\left(2\delta_2\left(\frac{1}{\tau_1^2}+\frac{1}{\tau_4^2}\right)J\tilde E^\dagger \tilde E J 
+ \frac{2}{\tau_3^2}J\tilde N^\dagger \tilde N J\right)P\nonumber \\
&&+ \frac{\tau_3^2+\tau_4^2}{2\gamma^2}\tilde E^\dagger \tilde E 
+ \frac{\tau_1^2}{2} \tilde N^\dagger \tilde N < 0. 
\end{eqnarray*}
Hence, we can choose $\tau_2 >0$ and $\tau_5 > 0$ sufficiently small so that 
\begin{eqnarray}
\label{QMI2}
\lefteqn{F^\dagger P + P F}\nonumber \\
&& +P\left(2\delta_2\left(\frac{1}{\tau_1^2}+\frac{1}{\tau_4^2}\right)J\tilde E^\dagger \tilde E J 
+ \frac{2}{\tau_3^2}J\tilde N^\dagger  \tilde N J\right)P\nonumber \\
&&+ \frac{\tau_3^2+\tau_4^2+\tau_5^2}{2\gamma^2}\tilde E^\dagger \tilde E 
+ \frac{\tau_1^2+\tau_2^2}{2} \tilde N^\dagger \tilde N < 0. 
\end{eqnarray}

Now, it follows from (\ref{z}) that we can write
\begin{eqnarray}
\label{zz}
\zeta^*\zeta &=& \left[\begin{array}{c}a \\ a^\#\end{array}\right]^\dagger \tilde E ^\dagger \tilde E \left[\begin{array}{c}a \\ a^\#\end{array}\right]. 
\end{eqnarray}
Also,  it follows from Lemma \ref{L2} that
\[
[V,\zeta] = -2 \tilde E 
JP\left[\begin{array}{c}a \\ a^\#\end{array}\right].
\]
Hence,
\begin{eqnarray}
\label{VzzV}
[V,\zeta][V,\zeta]^* =
4\left[\begin{array}{c}a \\ a^\#\end{array}\right]^\dagger PJ 
\tilde E^\dagger \tilde E
JP
\left[\begin{array}{c}a \\ a^\#\end{array}\right].
\end{eqnarray}
Similarly
\begin{eqnarray}
\label{VL1L1V}
[V,L_1][V,L_1]^* =
4\left[\begin{array}{c}a \\ a^\#\end{array}\right]^\dagger PJ 
\tilde N^\dagger \tilde N
JP
\left[\begin{array}{c}a \\ a^\#\end{array}\right].
\end{eqnarray}
In addition,
\begin{eqnarray}
\label{VL1L1V}
L_1^*L_1 &=& \left[\begin{array}{c}a \\ a^\#\end{array}\right]^\dagger \tilde N ^\dagger \tilde N \left[\begin{array}{c}a \\ a^\#\end{array}\right]. 
\end{eqnarray}

Hence using Lemma \ref{L2}, we obtain
\begin{eqnarray}
\label{lyap_ineq3}
&&-i[V,H]+ \mathcal{L}_{L_1}(V)
+\left(\frac{\tau_1^2}{2}+\frac{\tau_2^2}{2}\right)L_1^*L_1\nonumber \\
&&+\left(\frac{\delta_2}{2\tau_1^2}+\frac{\delta_2}{2\tau_4^2}\right)[V,\zeta]^*[V,\zeta]\nonumber \\
&&+\left(\frac{\tau_3^2}{2\gamma^2}+\frac{\tau_4^2}{2\gamma^2}+\frac{\tau_5^2}{2\gamma^2}\right)\zeta^* \zeta 
+\frac{[V,L_1]^*[V,L_1]}{2\tau_3^2}\nonumber \\
&=& \left[\begin{array}{c}a \\ a^\#\end{array}\right]^\dagger\left(\begin{array}{c}
F^\dagger P + P F\\ 
+2\delta_2\left(\frac{1}{\tau_1^2}+\frac{1}{\tau_4^2}\right)PJ\tilde E^\dagger \tilde E JP \\
+ \frac{2}{\tau_3^2}PJ\tilde N^\dagger \tilde NJP \\
+ \frac{\tau_3^2+\tau_4^2+\tau_5^2}{2\gamma^2}\tilde E^\dagger \tilde E \\
+ \frac{\tau_1^2+\tau_2^2}{2} \tilde N^\dagger \tilde N
\end{array}\right)\left[\begin{array}{c}a \\
a^\#\end{array}\right]\nonumber \\
&&+\tr\left(PJN^\dagger\left[\begin{array}{cc}I & 0 \\ 0 & 0 \end{array}\right]NJ\right)
\end{eqnarray}
where $F = -iJM-\frac{1}{2}JN^\dagger J N$. 

From this, it follows using (\ref{QMI2}) that there exists a constant $c > 0$ such that condition 
(\ref{dissip1a}) is satisfied with 
\[
\tilde \lambda = \tr\left(PJN^\dagger\left[\begin{array}{cc}I & 0 \\ 0 & 0 \end{array}\right]NJ\right) \geq 0.
\]
Hence, it follows from Lemma \ref{L4}, Lemma \ref{LB},  Theorem \ref{T2} and  $P > 0$ that 
\begin{eqnarray}
\label{ms_stable1}
\lefteqn{\left< \left[\begin{array}{c}a(t) \\ a^\#(t)\end{array}\right]^\dagger \left[\begin{array}{c}a(t) \\ a^\#(t)\end{array}\right] \right>}\nonumber \\
&\leq&  e^{-ct}\left< \left[\begin{array}{c}a(0) \\ a^\#(0)\end{array}\right]^\dagger \left[\begin{array}{c}a(0) \\ a^\#(0)\end{array}\right] \right>\frac{\lambda_{max}[P]}{\lambda_{min}[P]}\nonumber \\
&&+ \frac{\lambda}{c\lambda_{min}[P]}~~\forall t \geq 0
\end{eqnarray} 
where $\lambda = \tilde \lambda+ \left(\frac{\delta_3}{2\tau_2^2}+\frac{\delta_3}{2\tau_5^2}\right)\mu^*\mu
+\left( \frac{\tau_3^2}{2} + \frac{\tau_4^2}{2}+ \frac{\tau_5^2}{2}\right)\delta_1.$
 Hence, the condition (\ref{ms_stable0}) is satisfied with $c_1 = \frac{\lambda_{max}[P]}{\lambda_{min}[P]} > 0$, $c_2 = c > 0$ and $c_3 = \frac{\lambda}{c\lambda_{min}[P]} \geq 0$. 
\hfill $\Box$

\section{Conclusions}
\label{sec:conclusions}
In this paper, we have considered the problem of robust stability for
uncertain  quantum systems with  non-linear 
perturbations to the system coupling operator. The final stability result
obtained is expressed in terms of a strict bounded real 
condition. Future research will be directed towards analyzing the stability of specific nonlinear quantum systems
using the given robust stability result.


\bibliographystyle{IEEEtran}
\end{document}